\documentclass[%
 reprint,
 superscriptaddress,
 amsmath,amssymb,
 aps,
 floatfix,
]{revtex4-2}

\usepackage{graphicx}
\usepackage{dcolumn}
\usepackage{bm}

\usepackage[utf8]{inputenc}
\usepackage[T1]{fontenc}
\usepackage{mathptmx}
\usepackage{etoolbox}
\usepackage{CJKutf8}
\usepackage{booktabs}
\usepackage[version=3]{mhchem}
\usepackage[separate-uncertainty = true,multi-part-units=single]{siunitx}
\usepackage{listings}
\usepackage{csquotes}
\usepackage{gensymb}
\usepackage[dvipsnames]{xcolor}

\usepackage[colorlinks=true,
            linkcolor=blue,
            citecolor=blue,
            urlcolor=blue]{hyperref}

\DeclareSIUnit{\calorie}{cal}

\begin{document}

\title{From sparse quantum-computing data to atomistic simulation with universal machine-learning interatomic potentials}

\author{Tuan Minh Do}
\email{do.tuan.minh.qiqb@osaka-u.ac.jp}
\affiliation{Center for Quantum Information and Quantum Biology, The University of Osaka, 1-2 Machikaneyama, Toyonaka, Osaka 560-8531, Japan.}
\author{Yuichiro Yoshida}
\affiliation{Center for Quantum Information and Quantum Biology, The University of Osaka, 1-2 Machikaneyama, Toyonaka, Osaka 560-8531, Japan.}
\author{Kenji Ishihara}
\affiliation{Center for Quantum Information and Quantum Biology, The University of Osaka, 1-2 Machikaneyama, Toyonaka, Osaka 560-8531, Japan.}
\author{Wataru Mizukami}
\email{mizukami.wataru.qiqb@osaka-u.ac.jp}
\affiliation{Center for Quantum Information and Quantum Biology, The University of Osaka, 1-2 Machikaneyama, Toyonaka, Osaka 560-8531, Japan.}
\affiliation{Graduate School of Engineering Science, The University of Osaka, 1-3 Machikaneyama, Toyonaka, Osaka 560-8531, Japan.}

\date{\today}

\begin{abstract}
We propose a framework for incorporating quantum-computing-based electronic-structure calculations into universal machine-learning interatomic potentials (uMLIPs). Rather than constructing an interatomic potential from scratch, we refine a pretrained DFT-based uMLIP using a small set of accurate reference energies obtained from quantum computing. We demonstrate the approach for three chemically distinct applications: the Menshutkin reaction, water adsorption in the metal-organic framework HKUST-1, and CO hopping on a high-entropy-alloy nanoparticle. For the Menshutkin reaction, fine-tuning on gas-phase configurations improves the transition-state energy inside a carbon nanotube but not the product energy. For water adsorption in HKUST-1, fine-tuning with only 14 reference configurations brings adsorption thermodynamics obtained from millions of configurations sampled by Widom insertion into closer agreement with reference values. For CO hopping on an IrPdPtRhRu nanoparticle, the preference for on-top over bridge adsorption is recovered in the finite-temperature free-energy profile obtained from enhanced-sampling molecular dynamics, even though the reference data contain only energies. These results demonstrate that the proposed framework provides a practical route for incorporating quantum-computing calculations into realistic atomistic simulations and that quantum-computing reference data can improve pretrained uMLIPs.
\end{abstract}

\maketitle

\section{Introduction} \label{Introduction}
Quantum computing has advanced rapidly in recent years, and for some quantum many-body problems it is entering regimes where classical verification is becoming increasingly difficult~\cite{googleOTOC2025,andersenThermalization2025,quantinuumDigital2026,ibmDynamical2026}. One of the most promising applications of quantum computing is quantum chemistry~\cite{caoQuantumChemistryAge2019,mcardleQuantumComputationalChemistry2020}.
The application of quantum computing to electronic-structure calculations has developed substantially over the last decade.
It has been estimated that, once fault-tolerant quantum computers (FTQC) become available, the ground-state energy of the FeMo cofactor, a widely known system with a highly complex electronic structure, could be computed in less than a day~\cite{lowSpectralAmplification2025}.
As an intermediate step toward FTQC, researchers have also actively developed hybrid quantum--classical algorithms~\cite{tillyVariationalQuantumEigensolver2022,Jiang2025walking,Anurag2026towards}.
In particular, the emergence of sampling-based algorithms, represented by quantum-selected configuration interaction (QSCI)~\cite{kannoQuantumselectedConfigurationInteraction2026}, is beginning to enable practical quantum-chemistry calculations beyond toy-model systems~\cite{khinevichEnhancingQuantumComputations2025,erhartCoupledClusterMethod2024,erhartCoupledClusterMethod2025,yoshidaAuxiliaryfieldQuantumMonte2025,robledo-morenoChemistryScaleExact2025}.

However, quantum-computing calculations cannot yet provide the computational throughput required in modern computational chemistry.
Importantly, this limitation will not disappear simply with the advent of FTQCs. Rather, an FTQC necessarily involves substantial overhead from quantum error correction~\cite{babbushFocus2021}, and performing a single-point electronic-structure calculation can therefore remain time-consuming. 
At the same time, connecting electronic-structure accuracy to experimentally relevant finite-temperature properties requires accounting for contributions from many thermally accessible configurations. Atomistic simulation methods such as molecular dynamics provide a direct route to sampling these configurations but can require millions to billions of energy and force evaluations. 
This scale difference is reflected in recent studies that have explored directly combining quantum-computing-based electronic-structure methods with molecular dynamics~\cite{shiotaIntegratingClassicalQuantum2025,dasQuantumComputingEnabled2026}. A recently reported molecular-dynamics trajectory using quantum hardware was \SI{250}{fs}, corresponding to 500 time steps, which is far shorter than the trajectories typically required for extensive finite-temperature sampling.

The most direct approach to bridging this gap is to use electronic-structure calculations to construct an inexpensive potential energy surface (PES) model through machine learning. 
Such models are now commonly referred to as machine-learning interatomic potentials (MLIPs)~\cite{behlerGeneralizedNeuralNetworkRepresentation2007,deringerMachineLearningInteratomic2019,unkeMachineLearningForce2021,kocerNeuralNetworkPotentials2022,jacobsPracticalGuideMachine2025}, although the broader idea of constructing an inexpensive surrogate PES from expensive quantum-mechanical calculations has a much longer history, extending back to the early days of quantum mechanics. It is therefore natural to use expensive quantum-computing results to make an MLIP. 
Indeed, several studies have explored MLIPs trained on electronic-structure data obtained from quantum computers. 
However, so far, demonstrations have been restricted to systems with only a few atoms and to proof-of-concept simulations. 
Moreover, all these studies used the variational quantum eigensolver (VQE)~\cite{peruzzoVariationalEigenvalueSolver2014,tillyVariationalQuantumEigensolver2022} with small basis sets. Consequently, the resulting electronic-structure data were not only strongly affected by hardware noise but also lacked a substantial part of the dynamic electron correlation, resulting in not improving MLIPs. 
It therefore remains unclear whether combining quantum computing and MLIPs can provide a practical route to realistic atomistic simulation. Furthermore, previous studies have focused on constructing system-specific MLIPs from scratch. To our knowledge, the use of quantum-computing data to refine the recently emerging class of universal MLIPs (uMLIPs) has not yet been explored \cite{chenUniversalGraphDeep2022,dengCHGNetPretrainedUniversal2023,yangMatterSimDeepLearning2024,shiotaTamingMultiDomainFidelity2024,batatiaFoundationModelAtomistic2025}.

Here, we show that high-fidelity information from a limited number of quantum-computing-based electronic-structure calculations can be transferred to accurate downstream atomistic simulations through fine-tuning of pretrained uMLIPs. We demonstrate this across three chemically distinct applications using QSCI-based reference data \cite{doQuantumComputingAccurate2026}. The gas-phase energy profile for the Menshutkin reaction transfers to carbon-nanotube confinement and improves the predicted reaction barrier.
A static water-adsorption profile in HKUST-1 improves finite-temperature adsorption thermodynamics.
Finally, a static CO-hopping profile on an IrPdPtRhRu nanoparticle recovers the expected on-top site preference in the finite-temperature free-energy profile.

Our results show that the scarcity of quantum-computing reference calculations does not prevent their practical use in atomistic simulation. From a machine-learning perspective, they further suggest that quantum computers could provide a valuable source of high-fidelity training data for improving the capabilities of uMLIPs.

\section{Methods}

\subsection{Computational workflow}

\begin{figure*}
    \includegraphics{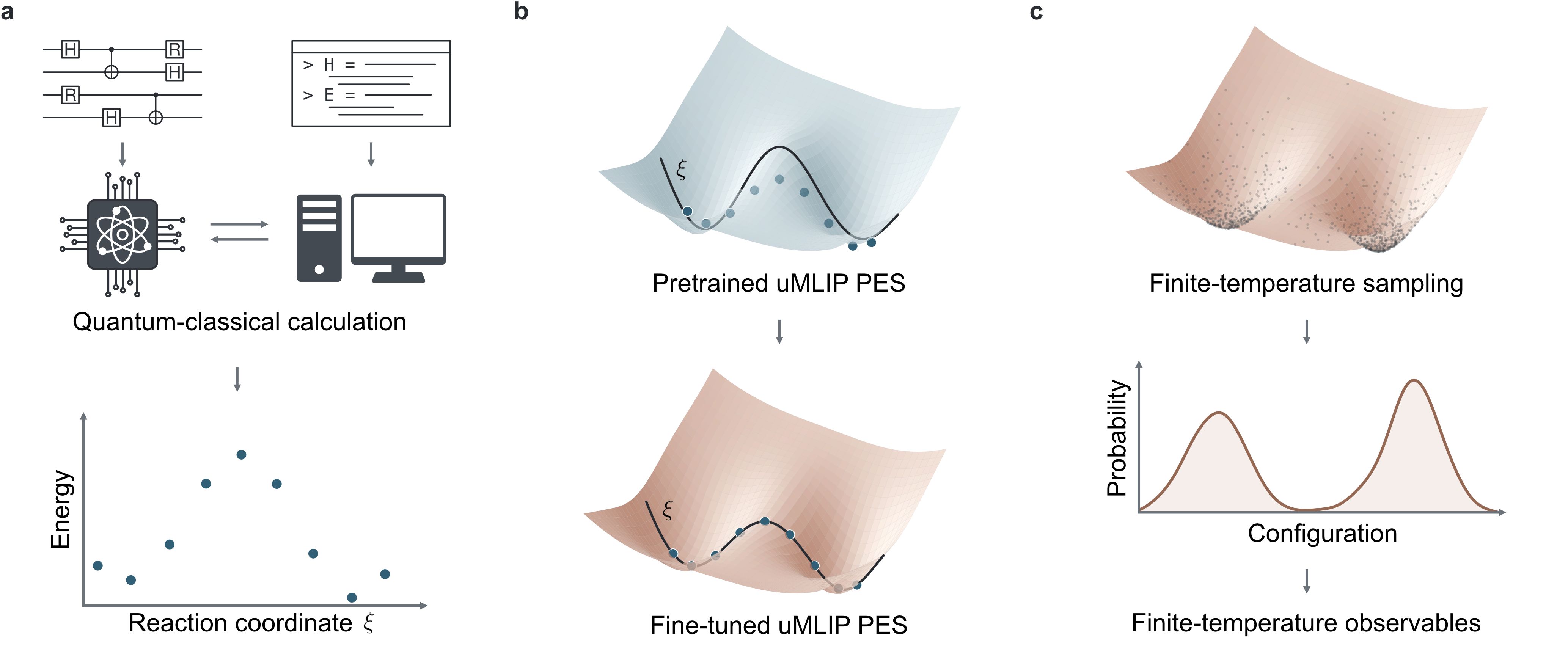}
    \centering
    \caption{\textbf{Computational workflow for transferring quantum-computing reference energies to finite-temperature atomistic simulations.}
    \textbf{a}, Quantum--classical electronic-structure calculations provide reference energies (blue points) for a small number of configurations along a reaction coordinate $\xi$.
    \textbf{b}, These reference energies are used to fine-tune a pretrained universal machine-learning interatomic potential (uMLIP). The energy profile along $\xi$ (black line) illustrates how the potential-energy surface (PES) changes after fine-tuning around the reference configurations.
    \textbf{c}, The fine-tuned uMLIP is subsequently used to sample an ensemble of configurations (gray points) at finite temperature. The resulting ensemble is used to calculate finite-temperature observables.
    }
    \label{fig:Workflow}
\end{figure*}

An overview of the computational workflow is shown in Fig.~\ref{fig:Workflow}.
High-fidelity energies are obtained from quantum-computing-based electronic-structure calculations.
Due to the high computational cost, only a small number of configurations can be evaluated in practice, for example as discrete points along a potential-energy curve.
The aim is to enable atomistic simulations to benefit from this level of accuracy despite the sparsity of the available high-fidelity information.

To achieve this, we fine-tune a pretrained uMLIP using the available data from the quantum-computing-based calculations. The pretrained uMLIP provides an approximate description of the potential-energy surface over a sufficiently broad region of configurational space to support stable atomistic simulations.
We exploit this information to interpolate between and extend beyond the configurations treated at the quantum-computing level, yielding an approximation to the corresponding potential-energy surface.

The resulting fine-tuned uMLIP can be evaluated efficiently throughout the extensive configurational sampling required for subsequent atomistic simulations.
The generated ensembles capture structural fluctuations at finite temperature and provide the statistical information needed to determine the relative populations of accessible configurations.
In this way, the workflow enables high-fidelity electronic-structure information from quantum computing to be transferred to finite-temperature atomistic simulations.

\subsection{Quantum-computing reference data}
The molecular structures and reference electronic energies used in this work were taken from our previous study \cite{doQuantumComputingAccurate2026}. 
The calculations in that study were based on a quantum--classical electronic-structure framework combining quantum-selected configuration interaction (QSCI) \cite{kannoQuantumselectedConfigurationInteraction2026}, projection-based wave-function-in-density-functional-theory (WF-in-DFT) embedding \cite{leeProjectionBasedWavefunctioninDFTEmbedding2019}, and classical post-processing.
Projection-based WF-in-DFT embedding partitions the full system into a wave-function subsystem and a surrounding environment described at the DFT level. Within the wave-function subsystem, QSCI provides an active-space wave function that is subsequently used in classical post-processing to recover electron correlation beyond the active space.

Tailored coupled-cluster theory with singles, doubles, and perturbative triples (TCCSD(T)) was used as the classical post-processing method for the reference energies considered here \cite{erhartCoupledClusterMethod2025}.
QSCI-TCCSD(T) was combined with WF-in-DFT embedding for water adsorption on the HKUST-1 MOF and CO hopping between bridge and on-top sites on the IrPdPtRhRu HEA nanoparticle. The gas-phase Menshutkin S$_\mathrm{N}$2 reaction was treated without embedding.
The QSCI calculations employed active spaces of (4e,4o), (2e,2o), and (4e,4o) for the Menshutkin, HKUST-1, and HEA systems, respectively, where the notation indicates the numbers of active electrons (e) and orbitals (o).
To obtain the determinants spanning the QSCI subspace, sampling in the computational basis was performed on the 144-physical-qubit superconducting quantum computer at the University of Osaka. Only the subset of qubits required by the corresponding active-space mapping was used, with a shot budget of 10,000 for each calculation.
The corresponding reference data contained 3, 14, and 9 configurations, respectively. The reference structures were used without further geometry optimization.

\subsection{MLIP fine-tuning}

Among the available MLIP architectures, we chose MACE \cite{batatiaMACEHigherOrder2022}, which combines equivariant message passing with many-body representations based on the atomic cluster expansion (ACE) \cite{drautzAtomicClusterExpansion2019}. Incorporating equivariance and explicit many-body representations can improve data efficiency \cite{batatiaMACEHigherOrder2022,batznerE3equivariantGraphNeural2022}. This matters here because only a small number of reference configurations is available. MACE-Osaka24 Large \cite{shiotaTamingMultiDomainFidelity2024}, a MACE foundation model trained on both molecular and inorganic crystal datasets, was used as the pretrained uMLIP for all calculations.
The reference data for the Menshutkin, HKUST-1, and HEA systems were combined into a single dataset and used to fine-tune one model using the multihead replay approach implemented in MACE \cite{tompaFinetuningMLIPFoundation2026}. 
In multihead replay, the target data and a replay dataset drawn from the original pretraining data are assigned separate output heads while sharing the underlying model representation.
This helps preserve information already learned by the pretrained uMLIP and limits its loss during fine-tuning, commonly known as catastrophic forgetting.
For the subsequent atomistic simulations, only the target head was used.

All available reference configurations were used for fine-tuning, with no subset held out for validation or testing. The reference data were therefore used for both training and validation of the target head. Instead of evaluating performance on a held-out subset of the reference data, the fine-tuned model was assessed using the subsequent atomistic simulations. No configurations from these simulations were included in the reference data.

For each system, the relative energies from the quantum-computing-based calculations were shifted by a constant offset such that the minimum-energy reference configuration was aligned with the corresponding energy predicted by the pretrained uMLIP.
The target reference data contained energy labels only, whereas the replay data included both energy and force labels.
Forces required for molecular dynamics and enhanced-sampling calculations were obtained as the negative gradient of the fine-tuned uMLIP energy with respect to the atomic coordinates.

Fine-tuning was performed using mace-torch version 0.3.15 with the Adam optimizer \cite{kingmaAdamMethodStochastic2015} and AMSGrad \cite{reddiConvergenceAdam2018} enabled, an initial learning rate of $1\times10^{-4}$, a batch size of 16, and a maximum of 301 epochs. The replay dataset comprised a randomly selected \SI{0.1}{\%} subset of the original pretraining data and was assigned a replay-head weight of 1.0. Global energy and force loss weights of 10 and 1 were used, respectively. Because the energy loss is normalized by the number of atoms, errors in the total energy contribute less to the loss for larger systems. To compensate for this system-size dependence, configuration-specific energy weights of 10, 100, and 100 were used for the Menshutkin, HKUST-1, and HEA reference data, respectively.

\subsection{Downstream atomistic simulations}

\subsubsection{Site-resolved Widom insertion calculations}

Water adsorption in HKUST-1 at infinite dilution was evaluated using site-resolved Widom test-particle insertion calculations~\cite{widomTopicsTheoryFluids1963} at \SI{295}{K}. 
The activated HKUST-1 structure was constructed from the crystal structure reported by Chui et al.~\cite{chuiChemicallyFunctionalizableNanoporous1999} by removing the Cu-bound and pore water molecules without further relaxation. 
The resulting cubic cell had a cell length of \SI{26.343}{\angstrom} and was treated as a rigid framework throughout the calculation.
The geometry of the inserted water molecule was taken from the QSCI-TCCSD(T)-in-DFT adsorption-energy calculations reported in Ref.~\cite{doQuantumComputingAccurate2026} and kept fixed during insertion.
For each calculation, $262\,144$ oxygen positions were sampled uniformly over the periodic cell. At each position, eight independent molecular orientations were sampled uniformly.

For each inserted configuration, the water--framework interaction energy $U_{\mathrm{int}}$ was calculated as
\begin{equation}
U_{\mathrm{int}} = E_{\mathrm{HKUST\text{-}1+H_2O}} - E_{\mathrm{HKUST\text{-}1}} - E_{\mathrm{H_2O}},
\end{equation}
where $E_{\mathrm{HKUST\text{-}1+H_2O}}$ is the energy of the framework containing the inserted water molecule, $E_{\mathrm{HKUST\text{-}1}}$ is the energy of the empty framework, and $E_{\mathrm{H_2O}}$ is the energy of the isolated water molecule. 
The minimum pair distances were defined as 0.57 times the sum of the corresponding van der Waals radii, following the default overlap criterion used in GROMACS~\cite{abrahamGROMACSHighPerformance2015}, and are listed in Table~\ref{tab:widom_cutoffs}.
Insertion attempts for which any water--framework atom pair was closer than the corresponding minimum distance were assigned zero Boltzmann weight while remaining included in the normalization of the Widom average.

\begin{table}[t]
    \centering
    \caption{Minimum pair distances used as cutoffs for Widom insertion.}
    \label{tab:widom_cutoffs}
    \begin{tabular}{lc}
        \toprule
        Pair & Minimum distance (\si{\angstrom}) \\
        \midrule
        C--H  & 1.65 \\
        C--O  & 1.84 \\
        Cu--H & 1.48 \\
        Cu--O & 1.66 \\
        H--H  & 1.37 \\
        H--O  & 1.55 \\
        O--O  & 1.73 \\
        \bottomrule
    \end{tabular}
\end{table}

To determine site-resolved adsorption properties, each sampled oxygen position was assigned to one of four mutually exclusive adsorption sites following the definitions of Ref.~\cite{garcia-perezIdentificationAdsorptionSites2009}.
Site~II comprised positions within \SI{2.0}{\angstrom} of the center of a small octahedral cage, whereas Site~III comprised positions between \SI{2.0}{\angstrom} and \SI{5.5}{\angstrom} from the same center.
Site~I was defined by a distance below \SI{3.0}{\angstrom} between the oxygen atom of the inserted water molecule and an open Cu site after excluding positions assigned to Sites~II and III. All remaining positions were assigned to Site~I$'$.

The adsorption enthalpy $\Delta H_s$ at infinite dilution for site $s$ was obtained from
\begin{equation}
\Delta H_s =
N_{\mathrm{A}}\frac{\left\langle U_{\mathrm{int}} \exp\left(-U_{\mathrm{int}}/k_{\mathrm{B}}T\right) \right\rangle_s}{\left\langle \exp\left(-U_{\mathrm{int}}/k_{\mathrm{B}}T\right) \right\rangle_s} -RT \,,
\end{equation}
where $\langle\cdots\rangle_s$ denotes the average over the sampled oxygen positions assigned to adsorption site $s$ and the eight molecular orientations sampled at each position. $N_{\mathrm{A}}$ is the Avogadro constant, $k_{\mathrm{B}}$ the Boltzmann constant, $R$ the molar gas constant, and $T$ the temperature. The difference in adsorption free energy $\Delta G_{\mathrm{I}}-\Delta G_{\mathrm{I'}}$ between Sites~I and I$'$ was calculated as
\begin{equation}
\Delta G_{\mathrm{I}}-\Delta G_{\mathrm{I'}} = -RT
\ln\left[\frac{\left\langle\exp\left(-U_{\mathrm{int}}/k_{\mathrm{B}}T\right)\right\rangle_{\mathrm{I}}}{\left\langle\exp\left(-U_{\mathrm{int}}/k_{\mathrm{B}}T\right)\right\rangle_{\mathrm{I'}}}\right] \,.
\end{equation}
The Henry coefficient $K_{\mathrm{H}}$ was determined from
\begin{equation}
K_{\mathrm{H}} =
\frac{V_{\mathrm{cell}}}{k_{\mathrm{B}}T M_{\mathrm{cell}}} \left\langle\exp\left(-\frac{U_{\mathrm{int}}}{k_{\mathrm{B}}T}\right) \right\rangle \,,
\end{equation}
where the average was taken over all sampled oxygen positions and molecular orientations, $V_{\mathrm{cell}}$ is the volume of the simulation cell, and $M_{\mathrm{cell}}$ is the molar mass of the framework.

The calculations were repeated three times. The same insertion positions and orientations were used for the pretrained and fine-tuned uMLIPs in each run. Thermodynamic quantities were evaluated separately for each run. Adsorption enthalpies and adsorption free-energy differences were then averaged across the runs, whereas Henry coefficients were averaged after taking their logarithms. Two-sided \SI{95}{\%} Student-$t$ confidence intervals were calculated from the three independent runs.

\subsubsection{CO hopping free-energy calculations}

The free-energy profile for CO hopping was calculated for a 79-atom IrPdPtRhRu high-entropy alloy (HEA) nanoparticle with composition Ir$_{16}$Pd$_{16}$Pt$_{16}$Rh$_{16}$Ru$_{15}$.
The path connected a bridge site between two neighboring Ir atoms and an on-top site on one of the two Ir atoms.
The system was modeled without periodic boundary conditions, and all atoms were allowed to move during the simulations.

A normalized path coordinate $\xi$ was defined from the carbon positions of the nine reference structures, with each position expressed relative to the midpoint between the two Ir atoms defining the bridge site.
A natural cubic spline was fitted through the corresponding relative carbon positions to define a continuous hopping path, with $\xi=0$ corresponding to the bridge site and $\xi=1$ to the on-top site.
During the simulations, the carbon position relative to the Ir--Ir midpoint was projected onto the spline, and the corresponding position along the spline defined the path coordinate $\xi$.
Overall rotation of the nanoparticle was suppressed by restraining its orientation using a harmonic potential with a force constant of \SI{2000}{\electronvolt\per\radian\squared}.
The nanoparticle orientation was determined from the positions of the 79 metal atoms.
The perpendicular distance $z$ of the carbon position from the spline was restrained using a flat-bottom harmonic potential,
\begin{equation}
U_z(z)=
\begin{cases}
0 \,, & z\leq z_0 \,,\\
\frac{1}{2}k_z(z-z_0)^2 \,, & z>z_0 \,,
\end{cases}
\end{equation}
with $z_0=\SI{0.15}{\angstrom}$ and $k_z=\SI{10}{\electronvolt\per\angstrom\squared}$.
No constraint was applied to the C--O bond.

Umbrella-sampling simulations \cite{torrieNonphysicalSamplingDistributions1977} were performed with OpenMM 8.5.2 \cite{eastmanOpenMM8Molecular2024} using the pretrained and fine-tuned uMLIPs through OpenMM-ML 1.7.
A harmonic umbrella potential
\begin{equation}
U_i(\xi)=\frac{1}{2}k_\xi(\xi-\xi_i)^2
\end{equation}
was applied in each window, where $\xi_i$ is the center of window $i$ and $k_\xi=36$~eV.
A total of 26 umbrella windows were used along the path coordinate.
NVT simulations were performed at \SI{300}{K} using Langevin dynamics with a friction coefficient of \SI{10}{ps^{-1}} and a timestep of \SI{0.5}{fs}.
Each window was equilibrated for \SI{2}{ps}, followed by a \SI{10}{ps} production run, with configurations recorded every \SI{5}{fs}.
The calculations were repeated five times.

Potential-of-mean-force (PMF) profiles were reconstructed separately for each run from the production trajectories using the weighted histogram analysis method (WHAM) \cite{kumarWeightedHistogramAnalysis1992} as implemented in Ref.~\cite{grossfieldWHAM}.
The path coordinate was divided into 130 bins from $-0.15$ to $1.15$, with a convergence tolerance of \SI{1e-9}{eV}.
The forward and reverse free-energy barriers were obtained from the difference between the PMF maximum and the values at the bridge ($\xi=0$) and on-top ($\xi=1$) endpoints, respectively.
Before averaging, each PMF was referenced to its minimum within the bridge region, defined as $-0.10\leq\xi\leq0.20$.
Mean PMFs and two-sided \SI{95}{\%} Student-$t$ confidence intervals were calculated from the five independent runs.
For visualization, the mean PMF and its confidence interval were shifted by a constant such that the minimum of the mean PMF was zero.

\section{Results and discussion}

\subsection{Menshutkin reaction inside a carbon nanotube}

\begin{figure}
    \includegraphics{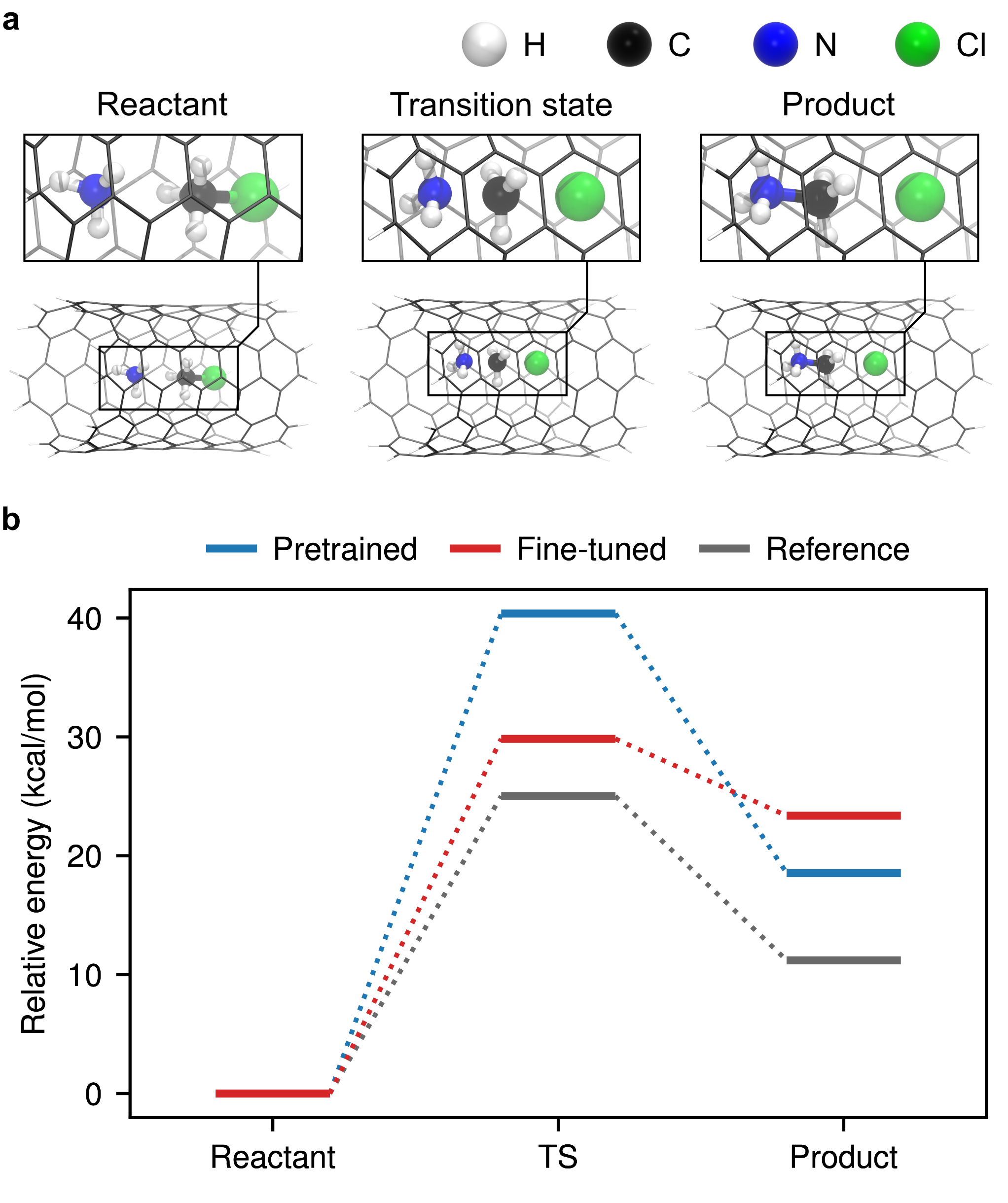}
    \centering
    \caption{\textbf{Transfer of gas-phase fine-tuning to the confined Menshutkin reaction.}
    \textbf{a}, Reactant, transition-state, and product structures of the Menshutkin reaction inside a carbon nanotube (CNT), with enlarged views of the reacting species.
    \textbf{b}, Relative energies of the reactant, transition state (TS), and product obtained with the pretrained and fine-tuned uMLIPs, together with THC-CASPT2 reference values from Ref.~\cite{songReducedScalingCASPT22018}.
    Energies are reported relative to the corresponding reactant.
    The uMLIP was fine-tuned on quantum-computing-based reference data for the gas-phase Menshutkin reaction and evaluated under confinement.}
    \label{fig:menshutkin_cnt}
\end{figure}

We first examine whether the correction learned from the gas-phase Menshutkin configurations transfers to the same reaction under confinement in a carbon nanotube (CNT).
The confined reactant, transition-state, and product structures were taken from Ref.~\cite{giacintoCNTConfinementEffectsMenshutkin2016}.
Single-point energy calculations were performed for the confined structures with the uMLIP before and after fine-tuning.
Reaction energies were reported relative to the confined reactant and compared with the corresponding THC-CASPT2 values from Ref.~\cite{songReducedScalingCASPT22018}.

The pretrained uMLIP substantially overestimates the barrier height of the confined reaction, predicting \SI{40.37}{\kilo\calorie\per\mole} compared with the THC-CASPT2 reference value of \SI{25.02}{\kilo\calorie\per\mole}.
After fine-tuning, the barrier decreases to \SI{29.83}{\kilo\calorie\per\mole}, reducing the absolute error from \SI{15.35}{\kilo\calorie\per\mole} to \SI{4.81}{\kilo\calorie\per\mole}.
In contrast, the product energy increases from \SI{18.54}{\kilo\calorie\per\mole} for the pretrained uMLIP to \SI{23.37}{\kilo\calorie\per\mole} after fine-tuning, increasing the deviation from the reference value of \SI{11.20}{\kilo\calorie\per\mole}.

As shown in Fig.~\ref{fig:S1}, the fine-tuned uMLIP closely reproduces the gas-phase reference energies used for fine-tuning, confirming that the gas-phase correction is learned successfully. 
The CNT results therefore show that the improvement achieved in the gas phase transfers only partially to the confined reaction, with the product energy becoming less accurate after fine-tuning. This suggests that including the target environment in the fine-tuning data may be important.
We therefore next consider two applications in which the quantum-computing-based calculations explicitly represent the environment relevant to the downstream simulations.

\subsection{Water adsorption thermodynamics in HKUST-1}

\begin{figure}
    \includegraphics[width=0.48\textwidth]{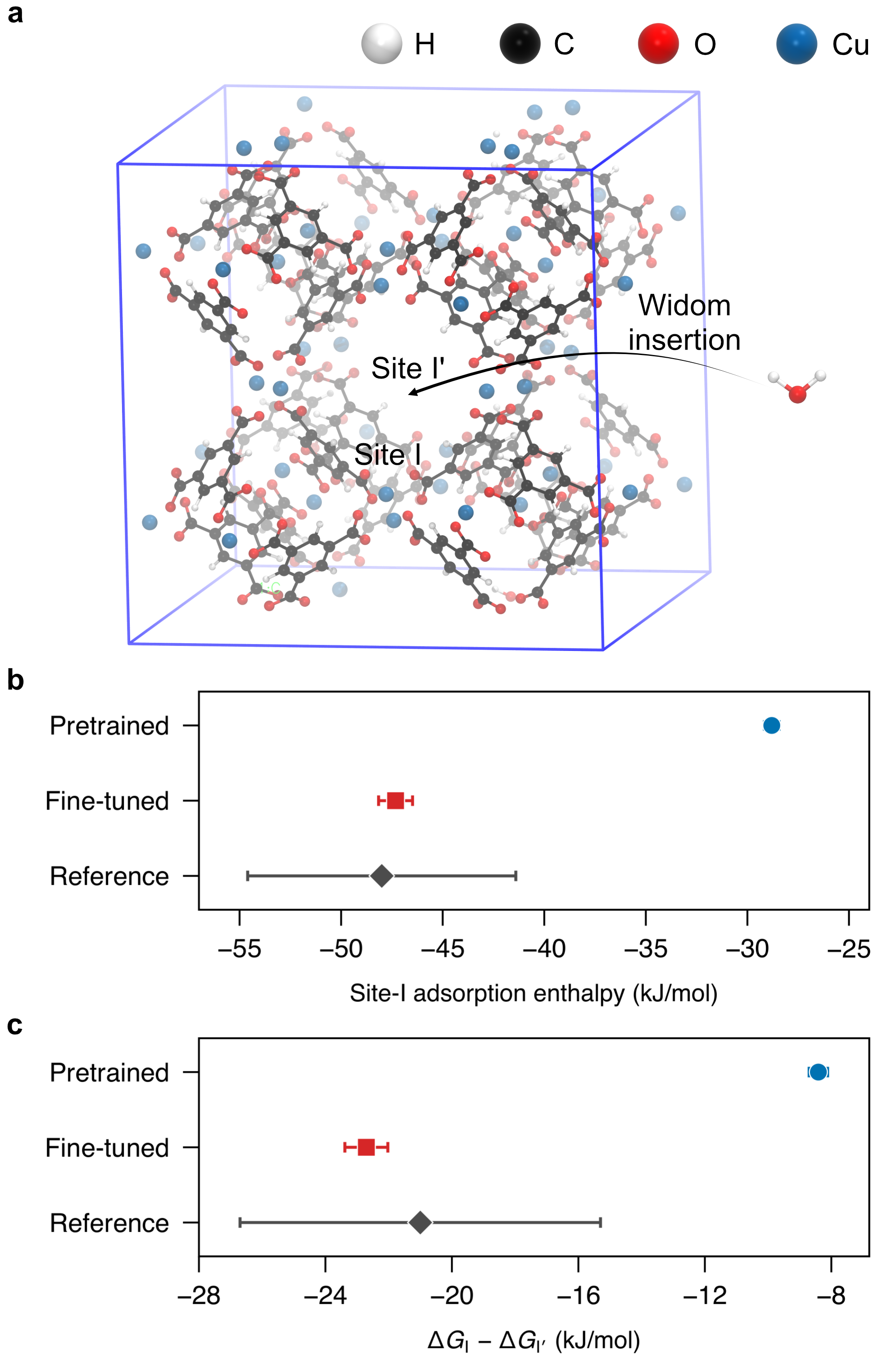}
    \centering
    \caption{\textbf{Transfer of quantum-computing-based adsorption energetics to adsorption thermodynamics in HKUST-1.}
    \textbf{a}, Widom insertion of water in HKUST-1, highlighting adsorption Sites~I and I$'$. The uMLIP is fine-tuned using quantum-computing-based reference energies along a water--Cu adsorption profile that covers the local environment of Site~I and also includes configurations associated with Site~I$'$.
    \textbf{b}, Site~I adsorption enthalpy at \SI{295}{K} obtained with the pretrained and fine-tuned uMLIPs, compared with the reference value reported in Ref.~\cite{castilloUnderstandingWaterAdsorption2008}.
    \textbf{c}, Free-energy difference between Sites~I and I$'$ at \SI{295}{K} obtained with the pretrained and fine-tuned uMLIPs and compared with the corresponding reference value from Ref.~\cite{castilloUnderstandingWaterAdsorption2008}.
    Error bars for the uMLIP results indicate \SI{95}{\%} confidence intervals from three independent calculations. Error bars for the reference values show the uncertainties reported in Ref.~\cite{castilloUnderstandingWaterAdsorption2008}.
    }
    \label{fig:hkust_widom}
\end{figure}

We next examine whether a water adsorption-energy profile at a Cu site of HKUST-1 obtained from quantum-computing-based calculations can be transferred through MLIP fine-tuning to accurately predict adsorption thermodynamics from Widom insertion calculations (Fig.~\ref{fig:hkust_widom}a). 
The adsorption-energy profile covers the local water--Cu environment of Site~I and also includes configurations associated with Site~I$'$.
We compare the adsorption enthalpy and free-energy difference calculated using the pretrained and fine-tuned uMLIPs with reference values reported in Ref.~\cite{castilloUnderstandingWaterAdsorption2008} (Fig.~\ref{fig:hkust_widom}b,c).

For Site~I, the pretrained uMLIP underestimates the strength of water adsorption. The adsorption enthalpy is \SI{-28.8}{\kilo\joule\per\mole}, whereas the reference value is \SI{-48.0}{\kilo\joule\per\mole}. 
Incorporating the quantum-computing-based adsorption profile through fine-tuning shifts the adsorption enthalpy to \SI{-47.3}{\kilo\joule\per\mole}, reducing the absolute deviation from \SI{19.2}{\kilo\joule\per\mole} to \SI{0.7}{\kilo\joule\per\mole}. Furthermore, the total adsorption enthalpy improves from \SI{-26.7}{\kilo\joule\per\mole} to \SI{-47.3}{\kilo\joule\per\mole}, compared with the reference value of \SI{-46.1}{\kilo\joule\per\mole}.

The transferred quantum-computing-based information also improves the relative thermodynamic stability of Sites~I and I$'$. The free-energy difference changes from \SI{-8.4}{\kilo\joule\per\mole} with the pretrained uMLIP to \SI{-22.7}{\kilo\joule\per\mole} after fine-tuning. The absolute deviation from the reference value decreases from \SI{12.6}{\kilo\joule\per\mole} to \SI{1.7}{\kilo\joule\per\mole}.

The improvement is more pronounced for the Henry coefficient, which depends exponentially on the adsorption free energy. The total Henry coefficient increases from \SI{2.0e-5}{\mole\per\kilogram\per\pascal} with the pretrained uMLIP to \SI{1.1e-2}{\mole\per\kilogram\per\pascal} after fine-tuning. Relative to the reference value of \SI{1.3e-2}{\mole\per\kilogram\per\pascal}, the pretrained uMLIP underestimates the Henry coefficient by nearly three orders of magnitude, whereas after fine-tuning the discrepancy is reduced to approximately \SI{20}{\%}. However, note that the reported uncertainty of the reference value is \SI{3.4e-2}{\mole\per\kilogram\per\pascal}. The comparison therefore mainly shows that fine-tuning brings the Henry coefficient into the correct order of magnitude.

These results show that quantum-computing-based information from a limited adsorption-energy profile can be transferred through a uMLIP to adsorption thermodynamics. The corresponding properties are obtained from more than two million insertion configurations in each Widom calculation, whereas only 14 static configurations were used for fine-tuning. The improvements in adsorption enthalpy, relative site free energy, and Henry coefficient thus indicate that the influence of the transferred information extends beyond the configurations provided for fine-tuning into the surrounding configurational space.

\subsection{Free-energy profile for CO hopping on an HEA nanoparticle}

\begin{figure}
    \includegraphics[width=0.48\textwidth]{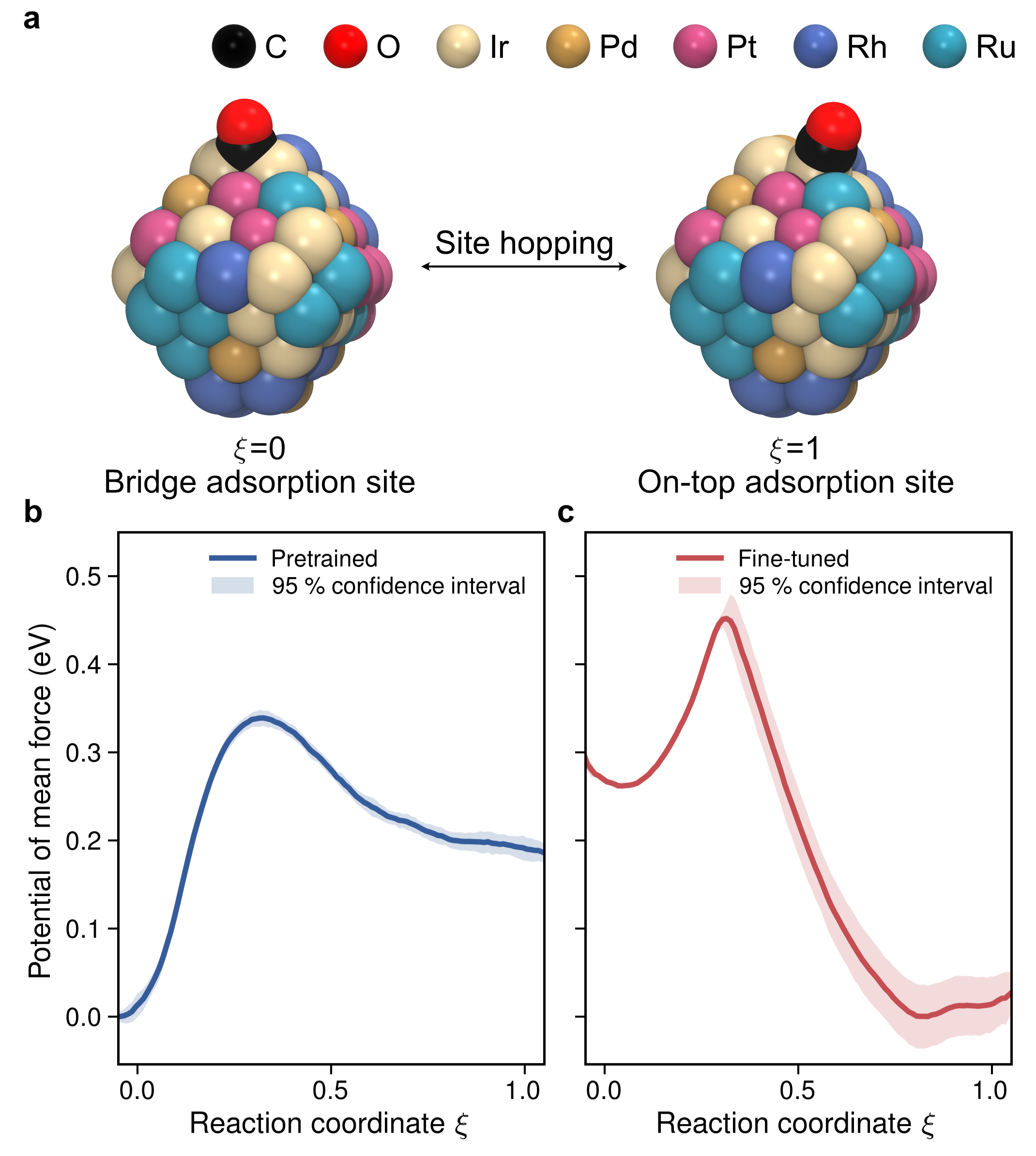}
    \centering
    \caption{\textbf{Transfer of quantum-computing reference data from a static hopping pathway to a finite-temperature free-energy profile on a high-entropy-alloy nanoparticle.}
    \textbf{a}, CO hopping from a bridge adsorption site ($\xi=0$) to an on-top adsorption site ($\xi=1$) on an IrPdPtRhRu high-entropy-alloy nanoparticle.
    \textbf{b}, Potential of mean force (PMF) at \SI{300}{K} obtained from umbrella-sampling simulations using the pretrained uMLIP.
    \textbf{c}, Corresponding PMF obtained using the uMLIP fine-tuned on the static QSCI-based hopping-energy profile.
    Fine-tuning reverses the relative stability of the two adsorption sites and changes the forward and reverse free-energy barriers.
    Solid lines show the mean PMFs over five independent runs, and shaded regions indicate the corresponding \SI{95}{\%} confidence intervals.}
    \label{fig:hea_pmf}
\end{figure}

Finally, we examine whether reference data from quantum-computing calculations along a static adsorbate-hopping pathway can be transferred to MD simulations to compute a free-energy profile. In addition to sampling configurations beyond the static reference set, such simulations require force evaluations to integrate the equations of motion, whereas the reference data provide only energies.
The reference profile describes CO hopping from a bridge site to an on-top site on an IrPdPtRhRu HEA nanoparticle (Fig.~\ref{fig:hea_pmf}a). To obtain the free-energy profile, we performed umbrella-sampling simulations at \SI{300}{K}. The corresponding PMFs for the pretrained and fine-tuned uMLIPs are shown in Figs.~\ref{fig:hea_pmf}b and c, respectively.

With the pretrained model, the forward free-energy barrier from the bridge ($\xi = 0$) to the on-top ($\xi = 1$) site is \SI{0.34}{eV}, while the reverse barrier is \SI{0.15}{eV}. The bridge site is favored over the on-top site by \SI{0.19}{eV}. After fine-tuning, the site preference is reversed, with the on-top site lying \SI{0.27}{eV} below the bridge site. The forward barrier decreases to \SI{0.16}{eV}, while the reverse barrier increases to \SI{0.43}{eV}.
For comparison, studies of CO adsorption on Ir(111) identify the on-top site as the preferred adsorption site \cite{krekelbergAtomicMolecularAdsorption2004,noeiMonitoringInteractionCO2018,liCOAdsorptionDisproportionation2022}. Experimental measurements on late-transition-metal surfaces report CO diffusion barriers of a few tenths of an electronvolt \cite{seebauerSurfaceDiffusionHydrogen1988,deckertSurfaceDiffusionCarbon1989,maDiffusionCOPt1111998}. The site preference after fine-tuning is consistent with these studies, and the forward and reverse barriers are similar in magnitude to the experimentally reported diffusion barriers.

These simulations demonstrate that a uMLIP fine-tuned on a small set of reference energies from quantum-computing-based calculations can be used for enhanced sampling at finite temperature. The reversal of the site preference together with the changes in the free-energy barriers shows that information from the static reference profile is transferred to the finite-temperature free-energy profile.

\section{Conclusions}

We have shown that a small amount of reference data from quantum-computing-based electronic-structure calculations can be incorporated into a pretrained uMLIP through fine-tuning to improve the accuracy of subsequent atomistic calculations and simulations.
A single uMLIP was jointly fine-tuned across three chemically distinct systems and evaluated upon transfer to a different chemical environment, as well as through Widom insertion calculations and in molecular dynamics simulations at finite temperature.

For the Menshutkin reaction, fine-tuning on gas-phase configurations reduced the error in the transition-state energy under CNT confinement but increased the error in the product energy, even though the gas-phase reference energies themselves were reproduced closely. This loss of accuracy may reflect partial forgetting of pretrained knowledge, which could be mitigated by optimizing the fine-tuning strategy and hyperparameters. Including the relevant chemical environment directly in the reference data may reduce sensitivity to the fine-tuning protocol. For the HKUST-1 and HEA applications, this was achieved using reference data from previous QSCI-based calculations in which the surrounding environment was treated through DFT embedding.

In HKUST-1, incorporating the QSCI-based reference data brought the thermodynamics of water adsorption derived from millions of configurations sampled by Widom insertion into closer agreement with reference values. This indicates that the corrections learned from the sparse reference data extend beyond the reference configurations into thermodynamically relevant regions of the configurational space. For CO hopping on the HEA nanoparticle, enhanced-sampling molecular dynamics yielded a free-energy profile that recovered the preference for the on-top over the bridge adsorption sites. This suggests that forces from quantum-computing-based calculations were not required during fine-tuning to obtain stable molecular dynamics trajectories at finite temperature and that energy-only reference data were sufficient to improve the resulting free-energy profile.

Our work demonstrates that quantum computers need not themselves perform all of the energy and force evaluations required for extensive finite-temperature atomistic simulations. Instead, they can be used to calculate highly accurate electronic energies for selected configurations. These energies can then be used to fine-tune a pretrained uMLIP, which serves as a surrogate for the much larger number of evaluations required during configurational sampling. This division of labor reduces the requirements for quantum computers to become practically useful in simulating complex molecular and materials systems. As quantum-computing-based electronic-structure methods continue to access more challenging problems, we anticipate that the proposed workflow will play an increasingly important role in bridging quantum computing and atomistic simulation.

\section{Acknowledgments}
This work was supported by MEXT Quantum Leap Flagship Program (MEXTQLEAP) Grant No. JPMXS0120319794 and the JST COI-NEXT Program Grant No. JPMJPF2014, and the JST ASPIRE Program Grant No. JPMJAP2319. 
A part of this work was supported by the Cross-ministerial Strategic Innovation Promotion Program (SIP), and JST Moonshot R\&D Grant No. JPMJMS256J.
We thank the Supercomputer Center, the Institute for Solid State Physics, the University of Tokyo, for allowing us to use their facilities. Part of the calculations were performed using the Genkai supercomputer of the Research Institute for Information Technology, Kyushu University, and the Miyabi supercomputer at the Information Technology Center, The University of Tokyo. This work was also achieved using the SQUID and OCTOPUS supercomputers at D3 Center, The University of Osaka.

\bibliography{Literature.bib}

\clearpage
\section{Supplemental information}

\setcounter{figure}{0}
\renewcommand{\thefigure}{S\arabic{figure}}

\begin{figure}[h!]
    \centering
    \includegraphics[width=0.4\textwidth]{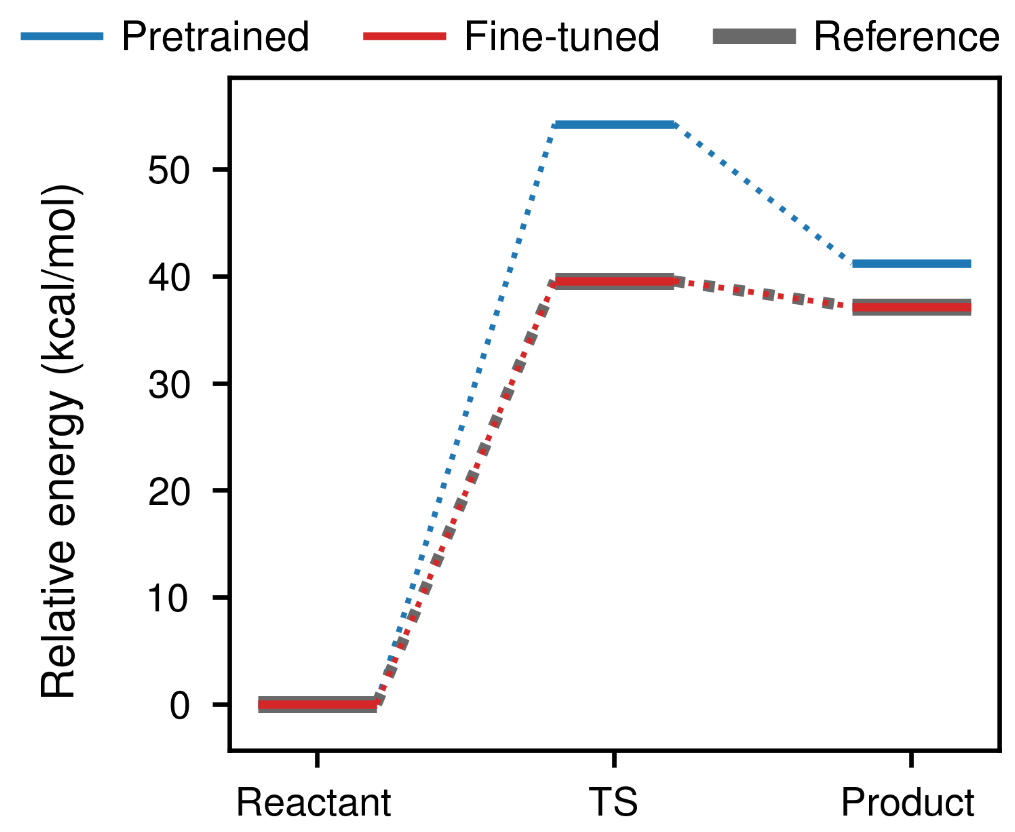}
    \caption{\textbf{Fine-tuning of the gas-phase Menshutkin reaction}.
    Relative energies of the reactant, transition state (TS), and product for the Menshutkin reaction obtained with the pretrained and fine-tuned uMLIP, together with QSCI-TCCSD(T) reference values from Ref.~\cite{doQuantumComputingAccurate2026}.}
    \label{fig:S1}
\end{figure}
\clearpage

\end{document}